# Normalized Weighting Schemes for Image Interpolation Algorithms

**Olivier Rukundo**

Clinical Physiology (Lund), Department of Clinical Sciences (Lund), Lund University, Skåne University Hospital, 221 85 Lund, Sweden. orukundo@gmail.com

**Abstract:** Image interpolation algorithms pervade many modern image processing and analysis applications. However, when their weighting schemes inefficiently generate very unrealistic estimates, they may negatively affect the performance of the end-user applications. Therefore, in this work, the author introduced four weighting schemes based on some geometric shapes for digital image interpolation operations. Moreover, the quantity used to express the extent of each shape's weight was the normalized area, especially when the sums of areas exceeded a unit square size. The introduced four weighting schemes are based on the minimum side-based diameter (MD) of a regular tetragon, hypotenuse-based radius (HR), the virtual pixel length-based height for the area of the triangle (AT), and the virtual pixel length for hypotenuse-based radius for the area of the circle (AC). At the smaller scaling ratio, the image interpolation algorithm based on the HR scheme scored the highest at 66.6% among non-traditional image interpolation algorithms presented. However, at the higher scaling ratio, the AC scheme-based image interpolation algorithm scored the highest at 66.6% among non-traditional algorithms presented, and, here, its image interpolation quality was generally superior or comparable to the quality of images interpolated by both non-traditional and traditional algorithms.

## 1. Introduction and Background

Commonly used normalization methods are based on interval arithmetic and fuzzy arithmetic [1]. Usually, normalization means rescaling variables in the range between zero and one [2]. While this meaning holds in this work, it can vary from problem to problem in other works [3],[4]. Here, the news is the normalization of areas whose sum exceeds a unit square size, especially when those areas are based on the Pythagorean theorem equation. In mathematics, the Pythagorean theorem is a fundamental relation in Euclidean geometry among the three sides of a right triangle [5]. Here, it can be written as Equation (1) or Equation (2) relating the lengths of the sides $a$, $b$, and $c$, of a right-angled triangle [5]. The Pythagorean equation has been widely used in many computer sciences and engineering techniques, including in [6], [7], but very few times for digital image interpolation purposes [8]. In mathematics, interpolation is an estimation method used to construct a new data value within the range of a set of known data [9]. In digital image processing, it is regarded as the process of generating estimates at each grid point which must preserve original pixels in the enlarged grid, because the image interpolation operation, by definition, must preserve the input pixel values of the smaller image. In real life, interpolation operation is routinely used for digital zooming [10]. Here, it is important to remember that digital zoom is one of the digital image processing techniques used to get a closer view of image objects or details of interest. Even if it is well known that digital zooming produces more visual artefacts than optical zooming in images, digital zooming is still widely used, for example, in modern digital cameras, imaging software (for processing and analysis applications in medicine, military, astronomy, etc.), as well as in many other electronic devices because, unlike optical zooming, digital zooming does not require the mechanical device of lens elements such as the one used in optical zoom [10]. To date, artefact-free digital image zooming remains very challenging to achieve due to unrealistic estimates by traditional interpolation algorithms—in addition to the current digital format requirements. Therefore, more research is still needed in this direction. Here, it is important to remember that interpolation pervades or penetrates many image processing and analysis applications. It is also important to remember that, in this work, developing novel and efficient weighting schemes was the goal because they could affect the performance of linear image resizing algorithms or desired results or the ways to obtain them [11]. To the best of the author's knowledge, there are currently many approaches to interpolation problems, particularly techniques developed to efficiently reduce visual artefacts in interpolated images or contribute to other image processing tasks [12-43]. Among them, some use deep learning means to reduce visual artefacts in the output of interpolated images [44],[45], which are beyond the scope of image interpolation methods presented, mainly because they do not respect the image interpolation definition [23], unlike those presented in this paper. One of my recent works, presented in [46], examined the origin of image pixels and divided image interpolation approaches into two major categories of non-extra-pixel and extra-pixel interpolation. Unlike the extra-pixel approach, the non-extra-pixel approach only depends on original or source image pixels [46]. Another of my recent works, presented in [10], examined the effects of rounding functions in one of the fastest image interpolation algorithms—the nearest neighbor interpolation algorithm. Of all my works on image interpolation, the most recently presented, in [47], examined the effects of stochastic rounding in nearest-neighbor interpolation for interpolated scan conversion operations in cardiac ultrasound. It is important to note that, unlike the non-extra pixel, the extra-pixel approach category has attracted many researchers to date due to the potential of leading to visually realistic image details after upscaling. In the extra pixel category, there are adaptive and non-adaptive interpolation techniques [8], [48], [49]. Moreover, among adaptive and non-adaptive, some techniques depend on traditional techniques, such as bilinear, to achieve improved outcomes relevant to the targeted problem or application of interest [48-55]. Here, the main contribution to the digital image processing field is the introduction of new weighting schemes for linear image interpolation algorithms inspired by the geometric shapes and related normalized areas when their sums exceed a unit square size. Here, it is important to remember that image interpolation algorithms belonging to the linear or non-adaptive category remain the most efficient among other interpolation algorithms. It is also important to remember that methods presented in this paper can be extended to 3D image interpolation operations to solve the specific problem of interest (in generating 3D models) by using them in simultaneous multidimensional interpolation operations [56], [57].

This paper is organized as follows: Part 1 presents the introduction and background. Part 2 presents the material and methods. Part 3 presents the results and discussions. The conclusion is presented in part 4.

## 2. Materials and Methods

### 2.1. *Pythagorean Theorem and Normalization*

Named after the Greek mathematician Pythagoras, the theorem can be reduced to Equation (1) and Equation (2).

$$a^2 + b^2 = c^2 \tag{1}$$

$$a^2 \leq c^2 - b^2 \quad (2)$$

where $a$ and $b$ are called legs or catheti of the right triangle and $c$ is called the hypotenuse, as shown in Figure 1.

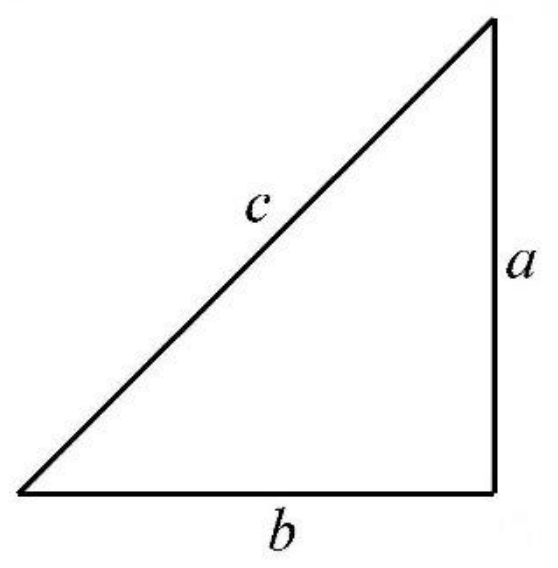

**Figure 1.** $c$ = hypotenuse, $b$ and a = legs (or base and height).

Now, referring to [2], the normalization of an $n$-tuple of general weights to the corresponding $n$-tuple of normalization weights is described by the real-vector-valued function $n: W_n \rightarrow S_n$ defined for all $(w_1, \ldots w_n) \in W_n$ in Equation (3). Here, it is important to remember that the normalization Equation (3) provides areas or weights whose sum is equal to one.

$$n(w_1, \ldots, w_n) := \left(\frac{w_1}{\sum_{i=1}^{n} w_i}, \ldots, \frac{w_n}{\sum_{i=1}^{n} w_i}\right) \quad (3)$$

*2.2. Normalized Weighting Schemes*

2.2.1. Tetragonal Area

A tetragon is another name for a quadrilateral, which is a geometric figure consisting of four sides or line edges, and which are connected by four corners or vertices [58],[59]. Note that a quadrilateral or tetragon has properties such as having four sides or edges, four vertices or corners, and interior angles that add to 360 degrees [58]. Figure 2 shows that a unit square (P1, P2, P3, P4) is composed of four tetragons with areas depending on the $x$-$y$ coordinates of $P$. Such areas, for example $W_3 = (x2 - x) \times (y - y1)$, are used as pixel weights and the $P$ value is obtained using Equation (4).

$$P_{(x,y)} = \sum_{i=1}^{4} (W_i \times P_i) \quad (4)$$

Here, it is important to remember that the sum of all weights must be equal to one.

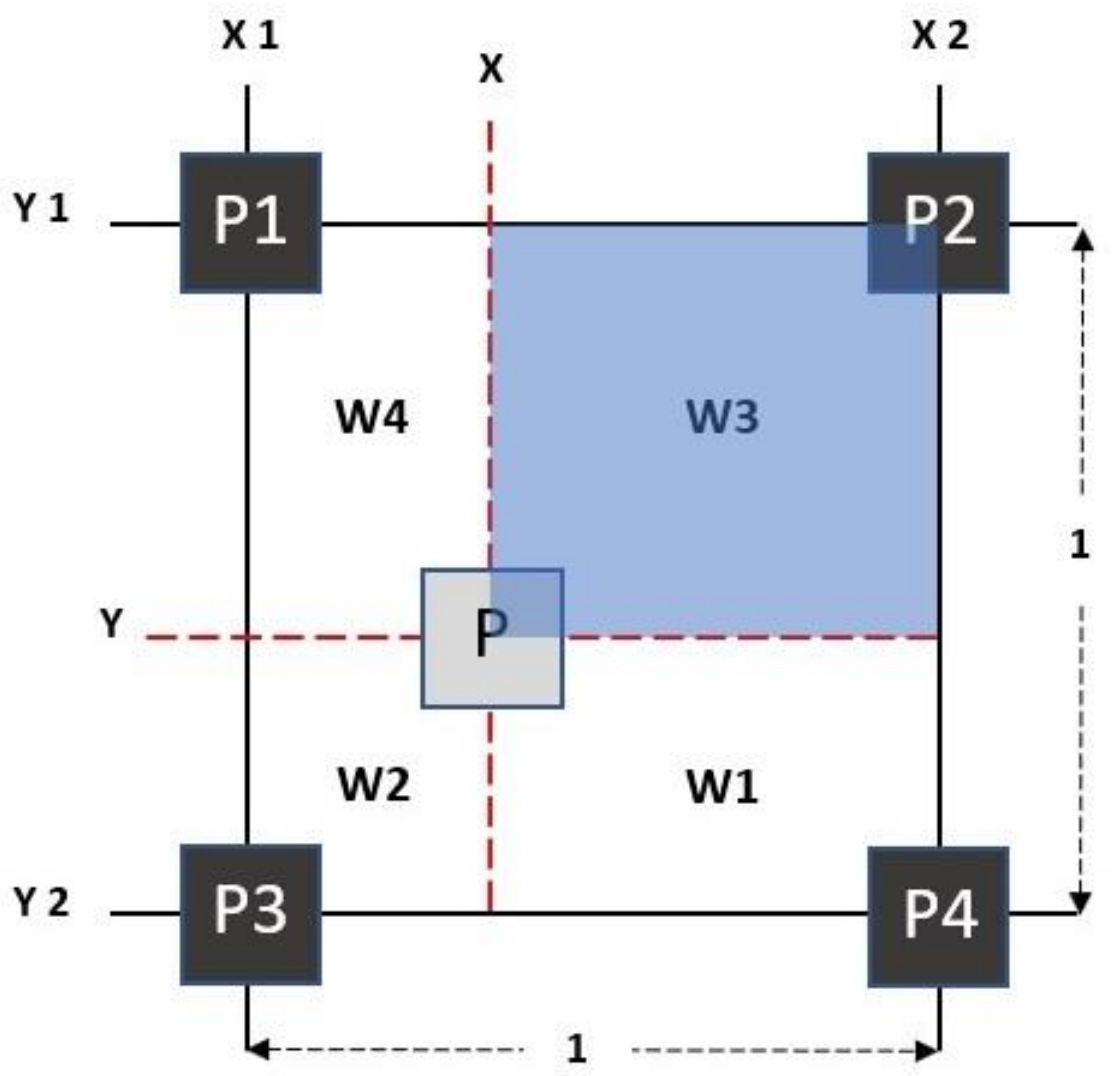

**Figure 2.** Example of a tetragon with four equal side lengths and angles—this is the same as the traditional bilinear (TB) weighting scheme.

Given that the areas of smaller tetragons, inside the main tetragon, sum up to one, in any case, similar to this, there will be no need for normalization of the sum of those areas. It is important to note that, throughout this paper, the amount of area is equivalent to the weight used to develop the weighting scheme.

2.2.2. Minimum Side-based Diameter

Here, the minimum side length, of any tetragon (located inside the main tetragon's unit square), is used as the diameter of a circle, as shown in Figure 3. Now, when the diameter is known, it becomes easy to find the area using Equation (5).

$$A = \frac{\pi}{4} \times diameter^2 \quad (5)$$

Here, it is important to note that the Equation (5)-based area replaces the weights ($W_i$) in Equation (4). However, given that the four circular areas do not sum up to one, in this case, there is a need for weight normalization using Equation (3).

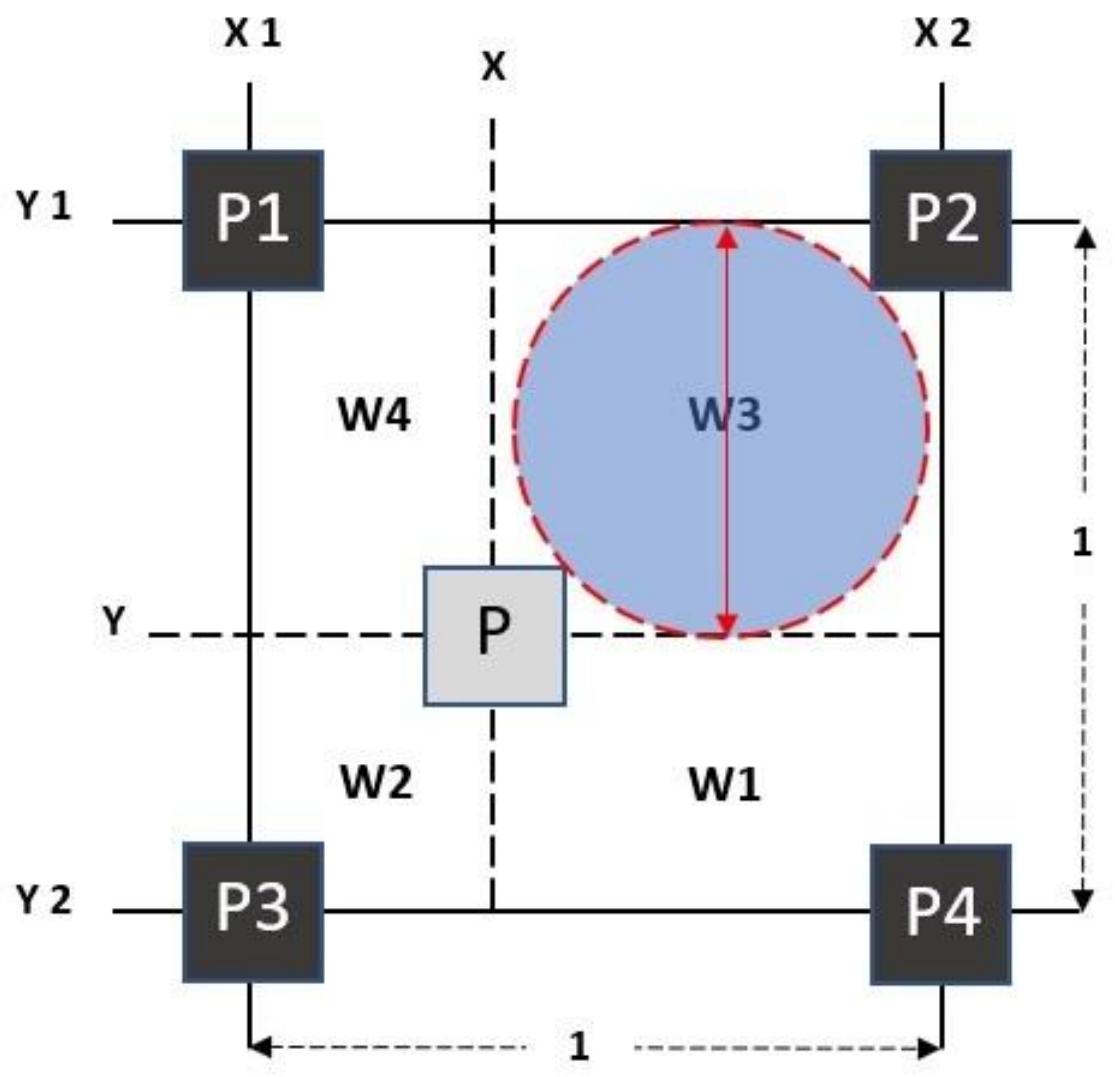

**Figure 3.** Example showing the minimum side diameter for the weighting scheme—this is the MD weighting scheme.

#### 2.2.3. Hypotenuse-based Radius

In the right-angled triangle, the hypotenuse is the longest side of the triangle. As shown in Figure 4, the hypotenuse is used as the radius of the circle. When the radius is known, the area of the circle is calculated using Equation (6).

$$A = \pi \times radius^2 \quad (6)$$

In addition, here, it is important to note that the Equation (6)-based area replaces the weights ($W_i$) in Equation (4). However, given that the four circular areas do not sum up to one, there is a need for weight normalization, using Equation (3).

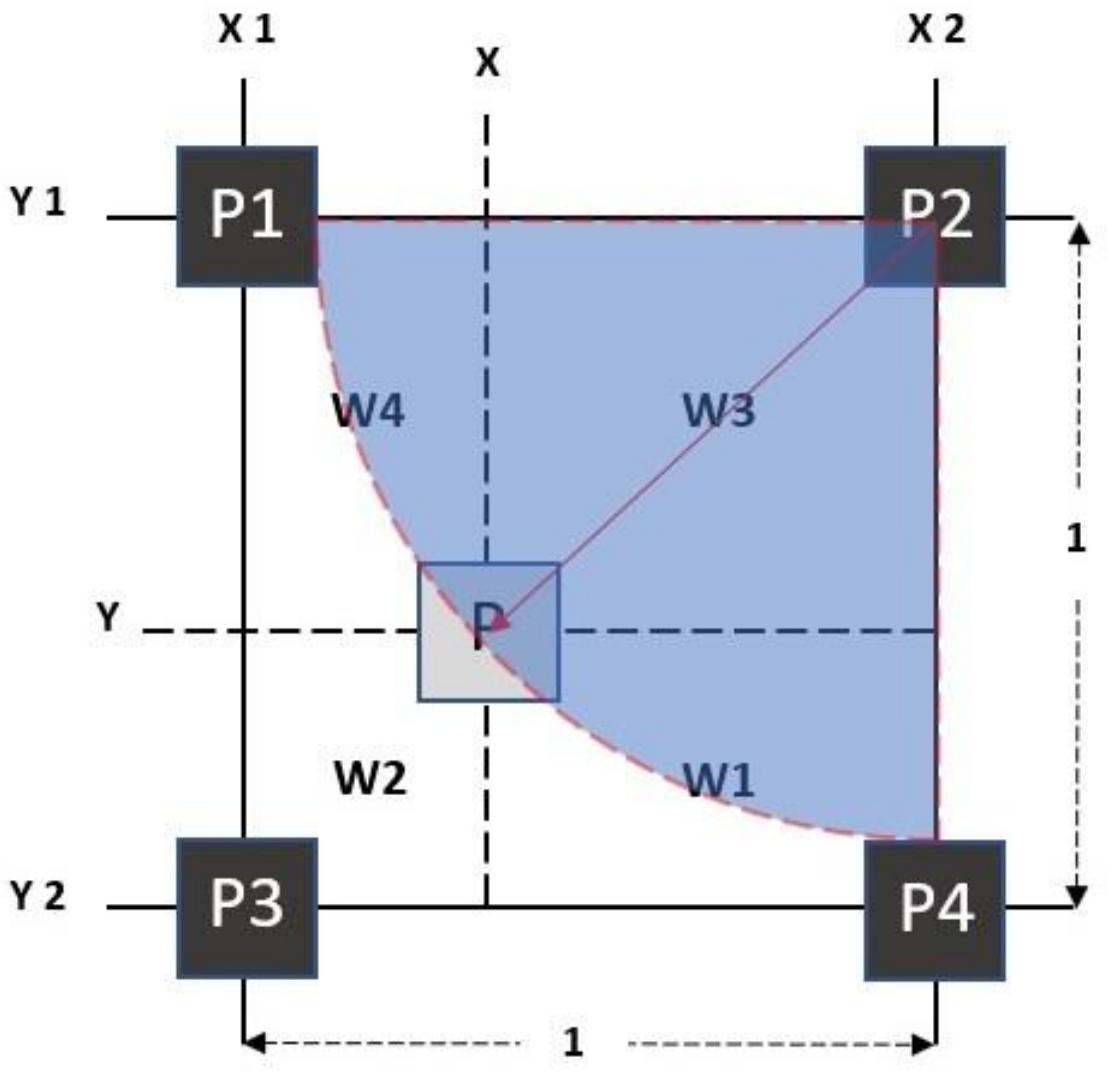

**Figure 4.** Example showing the hypotenuse-based radius for the weighting scheme – this is the HR weighting scheme.

#### 2.2.4. Preliminary Results

Figure 5 shows the preliminary interpolation results by the traditional bilinear or tetragon-based algorithm (TB), minimum side-based diameter algorithm (MD), and hypotenuse-based radius algorithm (HR). As can be seen, the basic structures and features, in the red square test image, were relatively recovered after image interpolation. Here, the scaling ratio was equal to 4. In this example, TB produced more blurriness artefacts than both MD and HR. However, HR produced more heavy jagged artefacts than both TB and MD. In all cases, HR produced a higher contrast image, with smoother edges, than both TB and HR algorithms cases. This means that the image interpolation quality-based performance, for each weighting scheme-based algorithm mentioned, is dependent on the spatial distribution of the input image.

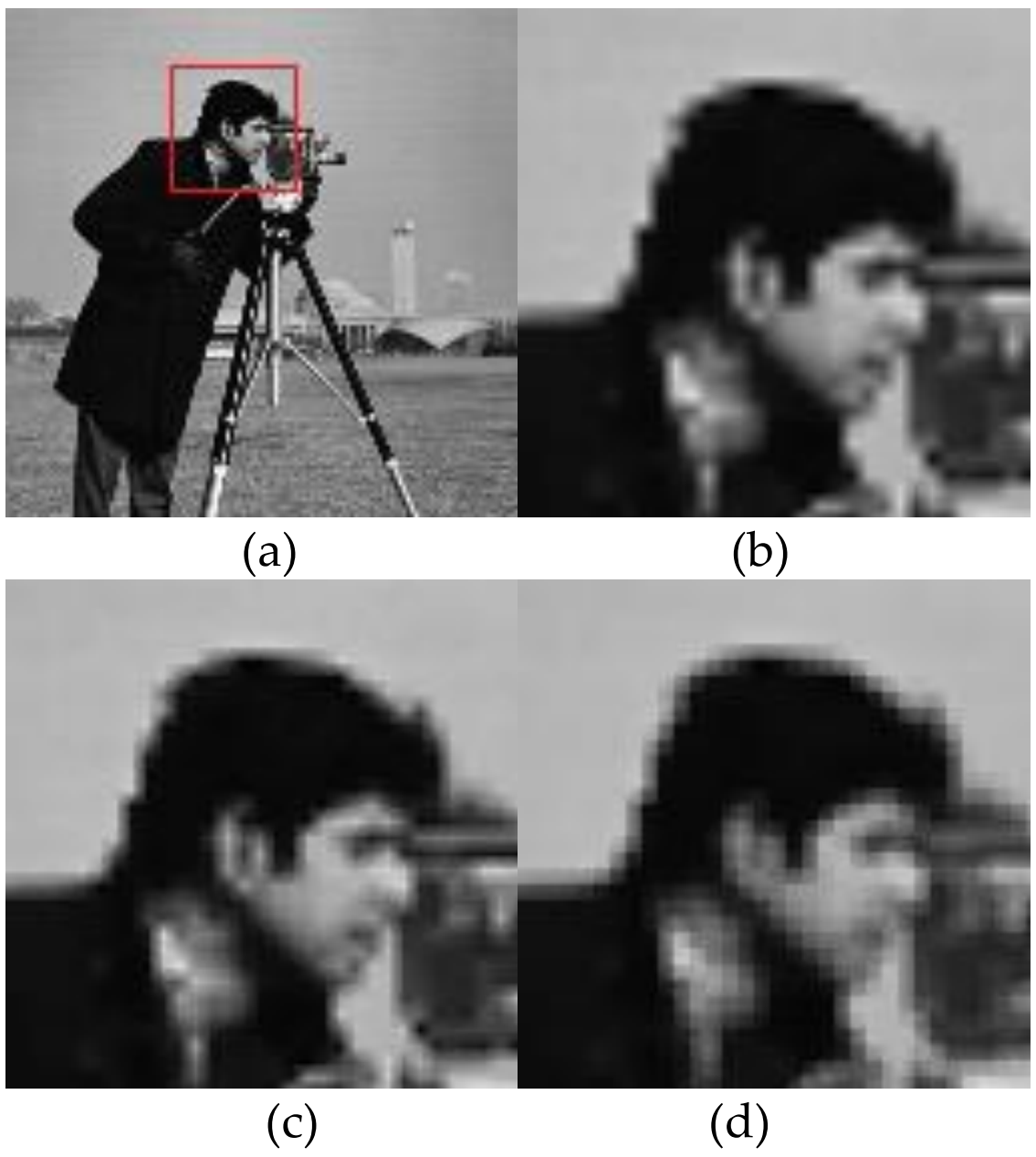

**Figure 5.** A test image-inside a red square (a) interpolated by the TB (b), MD (c), and HR (d).

*2.3. Virtual Pixel Length-based Normalized Weighting Schemes*

Here, the idea is that the virtual pixel length must be equal to the pixel value belonging to the pixel whose value belongs to the standard two-dimensional image grayscale range (which varies from 0 to 255 gray levels).

2.3.1. Virtual Pixel Length-based Height

In this scheme, the author referred to Figure 6, where the virtual pixel length ($A$) was used as the height of the triangle. Now, with the hypotenuse (B) as the base of the triangle, the area of the triangle can be calculated using Equation 7.

$$A = \frac{1}{2} \times Base \times Height \tag{7}$$

Here as in previous cases, it is important to note that the Equation (6)-based area replaces the pixel weights ($W_i$) in Equation (4). In addition, given that the four triangular areas do not sum up to one, there is a need for weight normalization using Equation (3).

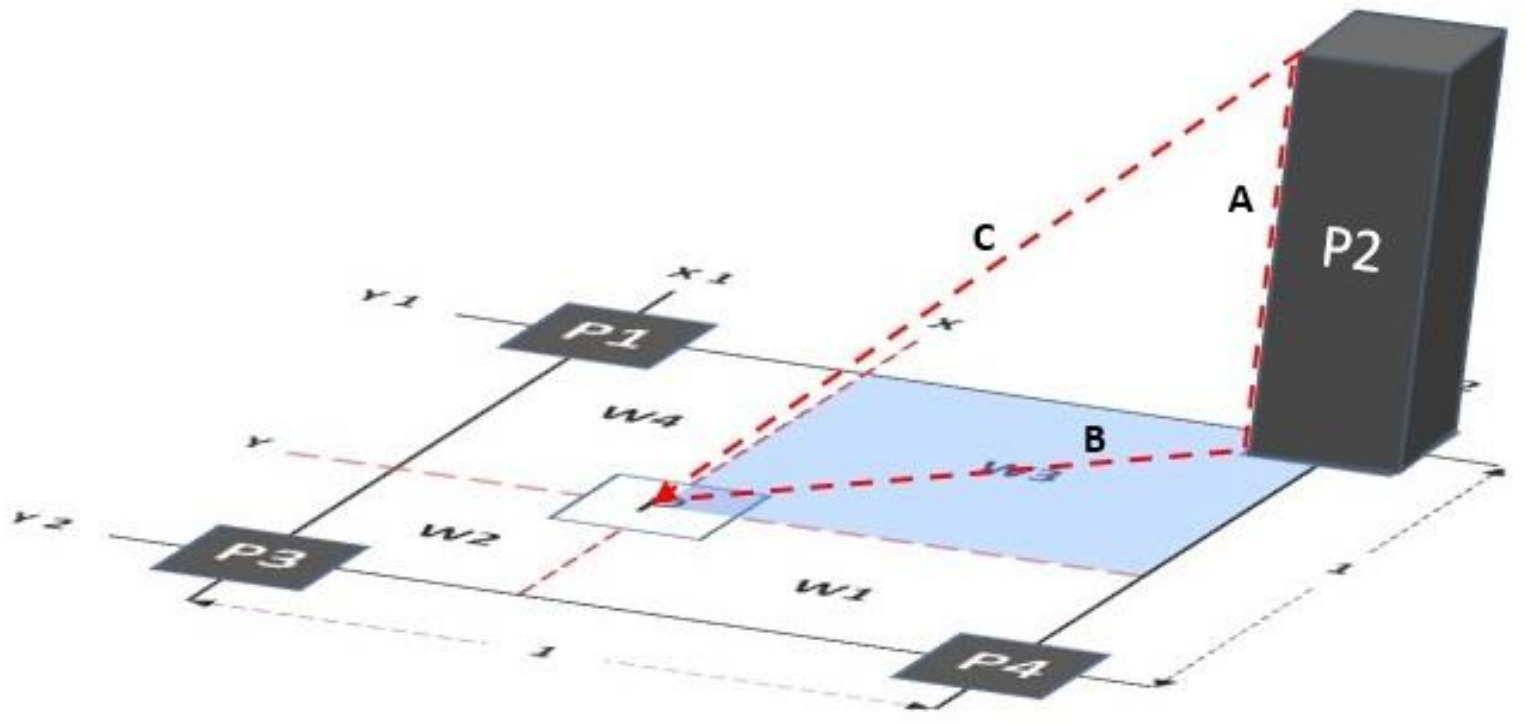

**Figure 6.** Example showing the virtual pixel length-based height for the area of the triangle weighting scheme—this is the AT weighting scheme.

2.3.2. Virtual Pixel Length for Hypotenuse-based Radius

Figure 7 shows the virtual pixel length (A), the base (B), and the hypotenuse (C). In this case, the hypotenuse (C) is used as the radius to find the area of a circle, using Equation (6).

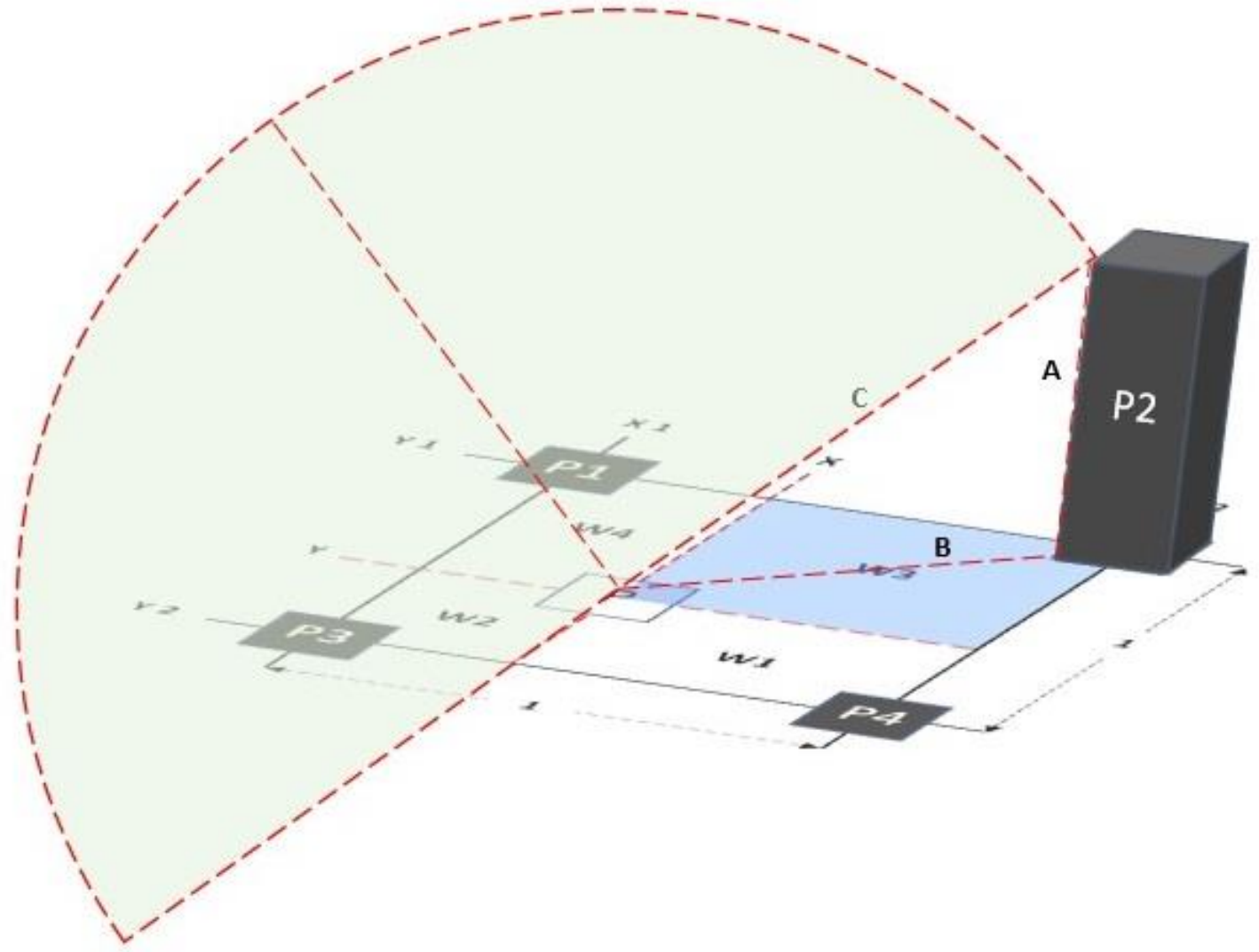

**Figure 7.** Example showing the virtual pixel length for hypotenuse-based radius for the area of the circle weighting scheme – this is the AC weighting scheme.

Referring to the example, shown in Figure 7, the Equation (6)-based area replaces the pixel weights, in Equation (4). Furthermore, given that the four circular areas do not sum up to one, there is a need for weight normalization, using Equation (3). Apart from that, it is important to note that this is inspired by one of the works presented in [28], in which the Pythagorean theorem equation was preliminary used to tackle the image interpolation algorithmic efficiency.

Figure 8 illustrates the simplified or generalized flowchart of the procedure of how the image interpolation generally works for upscaling/zooming-in purposes by TB, MD, HR, AT, and AC. Specifically, Figure 8 shows at which step interpolation weighting schemes are needed. Proposed image interpolation methods are different from traditional methods in terms of, or at the level of, the weighting schemes. Note that traditional nearest neighbor interpolation (TN) and traditional bicubic interpolation (TC) also follow the same procedure (as shown in Figure 8) but use different weighting schemes [10],[23]. In addition, note that, for image upscaling or zoom-in purposes, the value of ($x$) in Figure 8 must be at least equal to one.

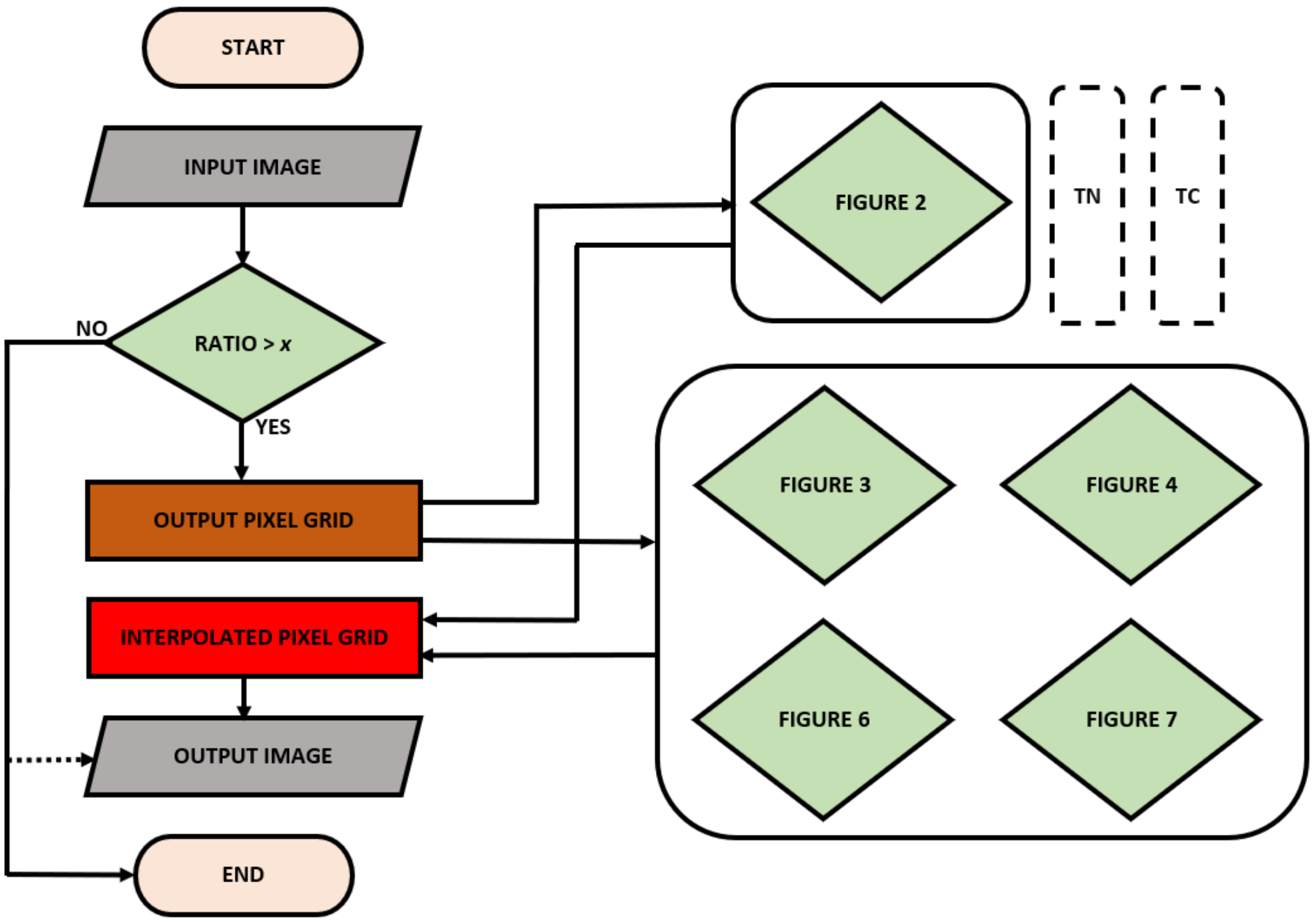

**Figure 8.** A simplified flowchart showing the image interpolation procedure followed by TB (Figure 2), MD (Figure 3), HR (Figure 4), AT (Figure 6), and AC (Figure 7) methods.

*2.4. Dataset*

In this work, the dataset consisted of 210 images of the category: textures, aerials, miscellaneous, and sequences were downloaded from the USC-SIPI Database [60]. Those images were converted to 8 bits and resized to 512 × 512, 256 × 256, and 128 ×128 using image processing functions available in the MATLAB software (R2020a). The resized versions are still accessible via the author's GitHub profile (see: GitHub.com/orukundo) [61].

*2.5. IQA Metrics*

Although there are two categories of objective image quality evaluation metrics, namely full-reference and non-reference metrics, widely used in evaluating image interpolation quality, in this work, only full reference (FR)-IQA metrics were chosen.

2.5.1. FR-IQA Metrics

Here, FR-IQA metrics were used due to the need for quantifying the closeness or similarity of interpolated images against their corresponding source or ground-truth images [49]. Those FR-IQA metrics selected included the mean-squared error (MSE), structural similarity index (SSIM), and peak signal-to-noise ratio (PSNR). The MSE is the mean square error between the test or input image and the ground truth or reference image. The PSNR is calculated using the following Equation (8), where K is the peak value either specified by the user or taken from the range of the image data type.

$$PSNR = 10 * log_{10}(\frac{K^2}{MSE}) \tag{8}$$

The SSIM is calculated based on the computation of 3 terms, namely: luminance (l), contrast (c), and structural (s) terms [62]. The resulting SSIM is a multiplicative combination of 3 terms, as shown in Equation (9), whose details can be found in [62].

$$SSIM(x, y) = [l(x, y)]^{\alpha} \times [c(x, y)]^{\beta} \times [s(x, y)]^{\gamma} \tag{9}$$

2.5.2. Speed Metrics

To evaluate the speed of each algorithm mentioned, MATLAB's TIC and TOC functions were used. Normally, the *tic* function works with the *toc* function to measure elapsed time. The *tic* function records the current time, and the *toc* function uses the recorded value to calculate the elapsed time. In our experiments, what was measured in reality was the source code line-reading speed of the mentioned image interpolation algorithms (at different scaling ratios).

## 3. Results and Discussions

This part presents results from objective and subjective evaluations or assessments of the quality of images interpolated by traditional and non-traditional image interpolation algorithms. The priority was given to the traditional algorithms such as the traditional nearest neighbor image interpolation algorithm (TN), traditional bicubic image interpolation algorithm (TC), and traditional bilinear image interpolation algorithm (TB). Those developed weighting schemes-based image interpolation algorithms (referred to as non-traditional image interpolation algorithms) to evaluate included: The normalized minimum side-based diameter (MD), normalized hypotenuse radius-based (HR), normalized virtual pixel length-based height for the area of the triangle (AT), and normalized virtual pixel length for hypotenuse-based radius for the area of the circle (AC) algorithms. Note that all these image interpolation algorithms were not only evaluated objectively and subjectively but also in terms of computational efficiency (using basic MATLAB functions to record the start and end time and print the time difference in seconds).

### *3.1. Objective Image Interpolation Quality Assessment*

In this section, the main goal was to use IQA metrics capable of predicting the quality of an interpolated image accurately and automatically. However, in Figure 9, the computational efficiency was evaluated by measuring the elapsed time. As can be seen, in Figure 9, the TN remained the fastest of all algorithms, because, in each scaling ratio case, it used the smallest average time, among the other algorithms mentioned. The elapsed time by the TB was shorter than MD, HR, AT, and AC. The main reason for the slowness of new or non-traditional algorithms was increased computational efforts which required more MATLAB lines reading time. For example, TN is the fastest because it does not do any computations. It only uses the destination image coordinates to locate the corresponding coordinates in the source image and those that match, their pixel values are copied from the source image to the destination image. That is not the case with any other algorithms mentioned, because, for example, the TB used the weighted average of four pixels to calculate each pixel value in the destination image. This required taking into account simultaneously a group of four pixels and sliding those four pixels kernel over the entire image, which understandably would take more time than a "copy-paste" like operation on which the TN was based. It is important to note that the software built-in interpolation functions of the traditional TN, TB, and TC normally perform better than the manually implemented functions.

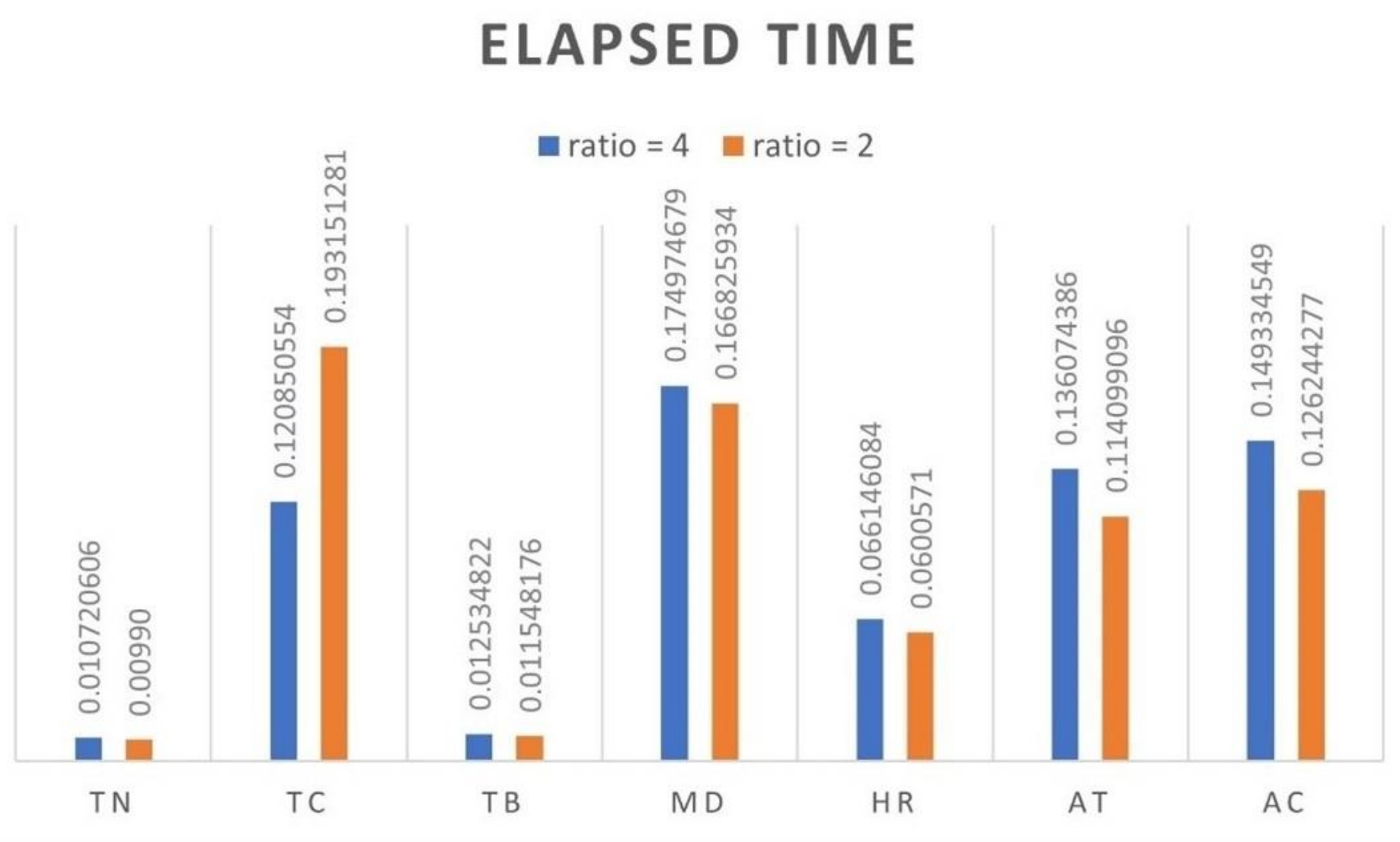

**Figure 9.** Average time (in seconds): traditional nearest neighbor (TN), traditional bicubic (TC), traditional bilinear (TB), normalized minimum side-based diameter-based (MD), normalized hypotenuse radius-based (HR), normalized virtual pixel length-based height for the area of the triangle-based (AT), and normalized virtual pixel length for hypotenuse-based radius for the area of the circle (AC) interpolation algorithms.

Now focusing on IQA metrics to predict automatically and/or accurately the image interpolation quality—Figure 10 presents the average scores that each interpolation algorithm produced or achieved in terms of the MSE. As can be seen, when the scaling ratio was 2 and 4, the TC achieved the best SME scores among traditional and non-traditional image interpolation algorithms. Focusing on non-traditional image interpolation algorithms alone (i.e., MD, HR, AT, and AC), it can be seen that the AC-based algorithm achieved the best MSE scores in both cases involving the scaling ratios equal to 2 and 4. Here, it is important to note that the lower MSE value (closest to zero) theoretically means the better image interpolation quality.

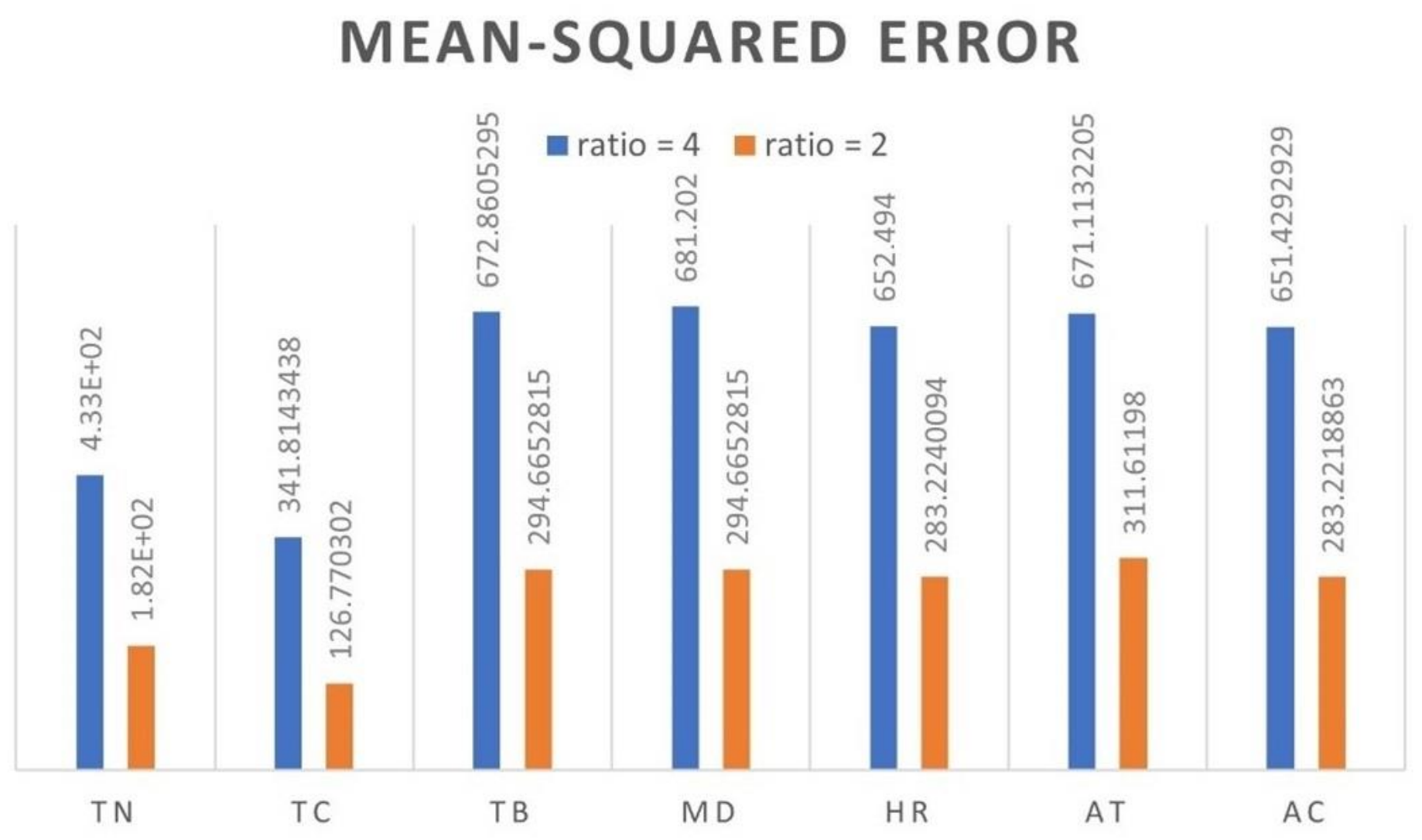

**Figure 10.** Average MSE: traditional nearest neighbor (TN), traditional bicubic (TC), traditional bilinear (TB), normalized minimum side-based diameter-based (MD), normalized hypotenuse radius-based (HR), normalized virtual pixel length-based height for the area of the triangle-based (AT), and normalized virtual pixel length for hypotenuse-based radius for the area of the circle (AC) interpolation algorithms.

Figure 11 presents the average scores, each interpolation algorithm produced or achieved in terms of the SSIM. As can be seen, for the cases involving all image interpolation algorithms as well as the scaling ratio equal to 2 and 4, TC achieved the best SSIM scores among all image interpolation algorithms mentioned. Focusing on image interpolation algorithms based on new weighting schemes alone, when the scaling ratio was 2, the HR-based algorithm achieved better SSIM scores than AC, AT, and MD-based image interpolation algorithms. In addition, when the scaling ratio was 4, the MD-based algorithm achieved better SSIM scores than AC, AT, and HR-based image interpolation algorithms. It is important to remember that, in theory, the higher SSIM score (closest to 1), means the better image interpolation quality.

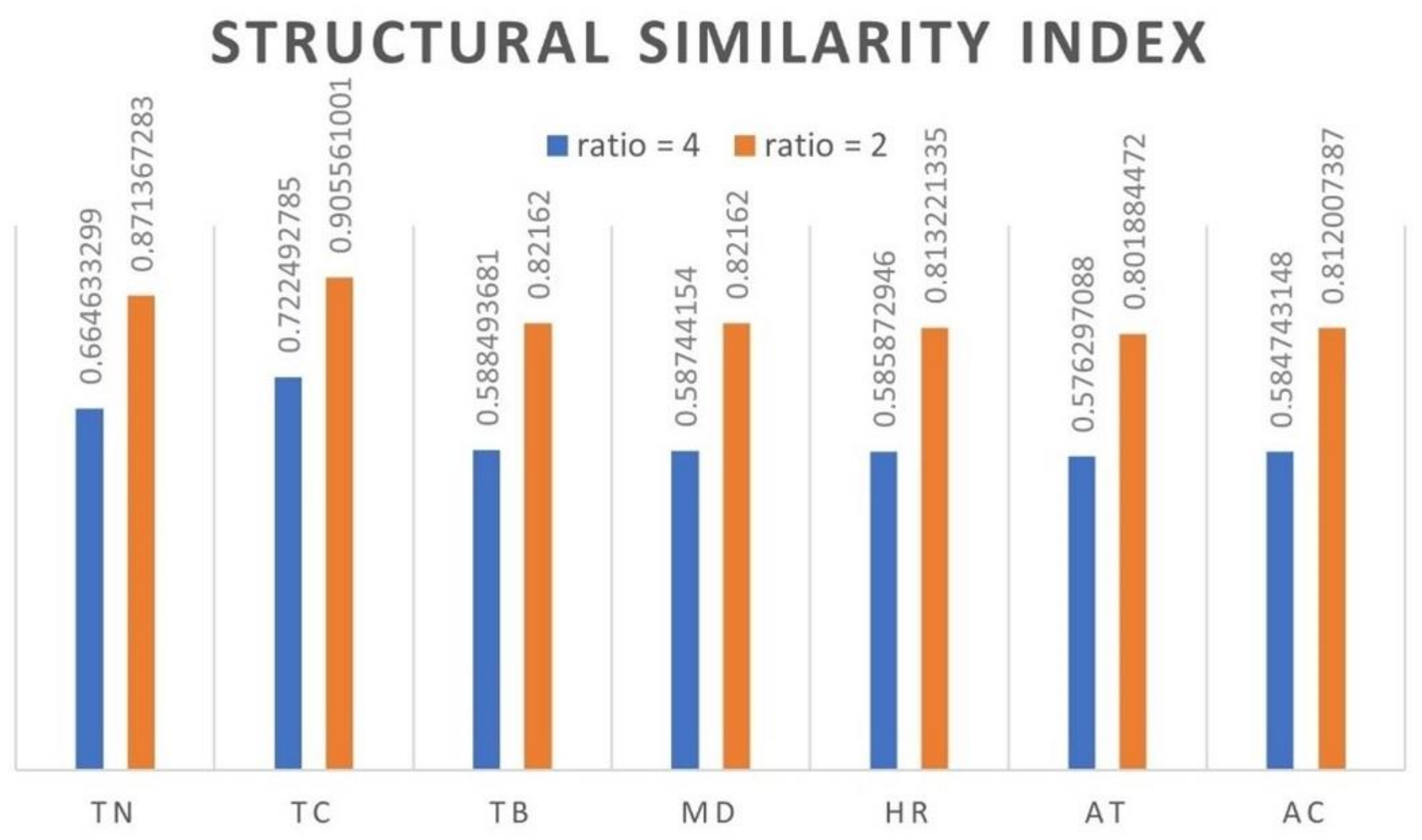

**Figure 11.** Average SSIM: traditional nearest neighbor (TN), traditional bicubic (TC), traditional bilinear (TB), normalized minimum side-based diameter-based (MD), normalized hypotenuse radius-based (HR), normalized virtual pixel length-based height for the area of the triangle-based

(AT), and normalized virtual pixel length for hypotenuse-based radius for the area of the circle (AC) interpolation algorithms.

Figure 12 presents the average scores, each interpolation algorithm produced or achieved in terms of the PSNR. Among all image interpolation algorithm cases, TC achieved the best PSNR scores. However, among non-traditional image interpolation algorithms, when the scaling ratio was 2, the HR-based algorithm achieved better PSNR scores than AC, AT, and MD-based image interpolation algorithms. Moreover, when the scaling ratio was 4, the AC-based algorithm achieved better PSNR scores than MD, AT, and HR-based image interpolation algorithms. In addition, here, it is important to remember that, in theory, the higher PSNR score (closest to 89) means the better image interpolation quality.

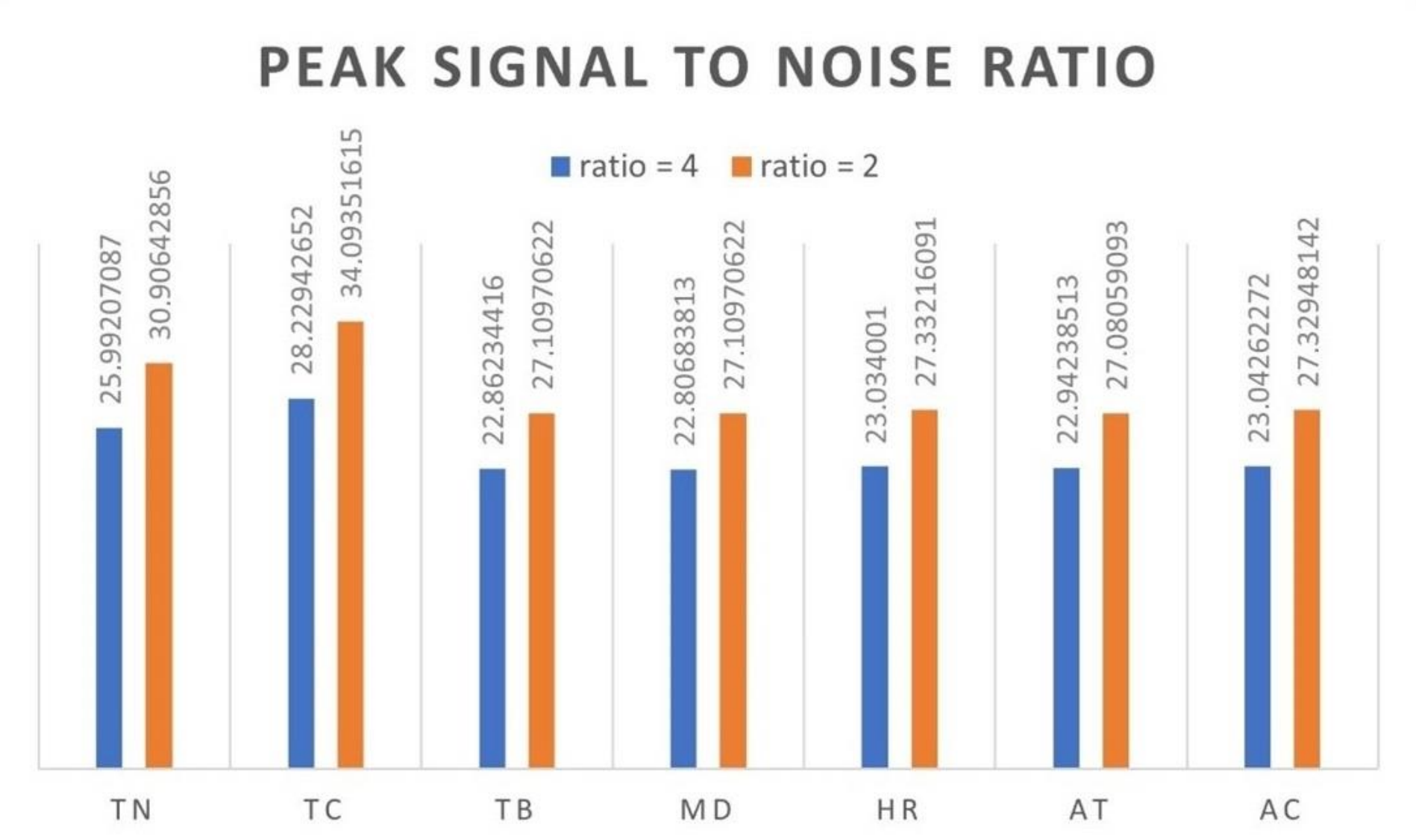

**Figure 12.** Average PSNR: traditional nearest neighbor (TN), traditional bicubic (TC), traditional bilinear (TB), normalized minimum side-based diameter-based (MD), normalized hypotenuse radius-based (HR), normalized virtual pixel length-based height for the area of the triangle-based (AT), and normalized virtual pixel length for hypotenuse-based radius for the area of the circle (AC) interpolation algorithms.

To simplify the interpretation of results presented in Figure 10, Figure 11, and Figure 12, a simplified representation or interpretation scheme of results (from both non-traditional algorithms or new schemes-based algorithms and traditional algorithms) was given or presented in Table 1 and Table 2. In these two tables, the maximum number of times each image interpolation algorithm was expected to score highest was equal to three because there were three IQA metrics, namely: MSE, SSIM, and PSNR. Here, a method that outperformed all others, in terms of the above three metrics scores, would score the highest score at 100%. For instance, if in each case of MSE, SSIM, and PSNR metrics, the TN achieved the highest scores, that means that TN achieved the highest score 3 times. In other words, TN achieved the highest at 100% because 3 times was the maximum number of times each image interpolation algorithm was expected to achieve the highest scores.

**Table 1.** Simplified representation of results (traditional vs. non-traditional).

| | TN | TB | TC | MD | HR | AT | AC |
|---|---|---|---|---|---|---|---|
| 2 x | 0 | 0 | 3 | 0 | 0 | 0 | 0 |
| 4 x | 0 | 0 | 3 | 0 | 0 | 0 | 0 |

**Table 2.** Simplified representation of results (non-traditional).

| | MD | HR | AT | AC |
|---|---|---|---|---|
| 2 x | 0 | 2 | 1 | 1 |
| 4 x | 1 | 0 | 0 | 2 |

Now, referring to the above example, it can be seen that, when the scaling ratio was equal to 2 and 4, TC achieved the highest at 100% among all image interpolation algorithms mentioned. However, among non-traditional image interpolation algorithms alone, when the scaling ratio was 2, the HR-based image interpolation algorithm achieved the highest at 66.6%. However, when the scaling ratio was equal to 4, the AC-based image interpolation algorithm achieved the highest 66.6% among all non-traditional image interpolation algorithms.

*3.2. Subjective Image Interpolation Quality Assessment*

In this section, we did subjective evaluations, which, in this context, was the most accurate and reliable way of evaluating the image interpolation quality. However, since they were generally expensive, external-factor dependent (including the observer's mood), and time-consuming, and they required a human observer (i.e., the author); as a result, they were limited to a few but most representative image examples shown in Figure 13 (the boat), Figure 14 (the man), and Figure 15 (the clock). As can be seen, in Figure 13, Figure 14, and Figure 15, skipping the image with a red square and considering the direction from left to right, there are interpolated images by TN, TC, and TB interpolation algorithms in the first row. In the second row, there are MD, HR, AT and AC interpolated images. Focusing on Figure 13, it can be seen that some edges are jagged while others are not, especially the vertical and horizontal edges of the TN image. Here, the TN algorithm did not produce blurred edges. In the TC image, the oblique, vertical, and horizontal edges look blurred to some extent but not jagged. In the TB image, regardless of the edge direction, the edges are blurred but not jagged.

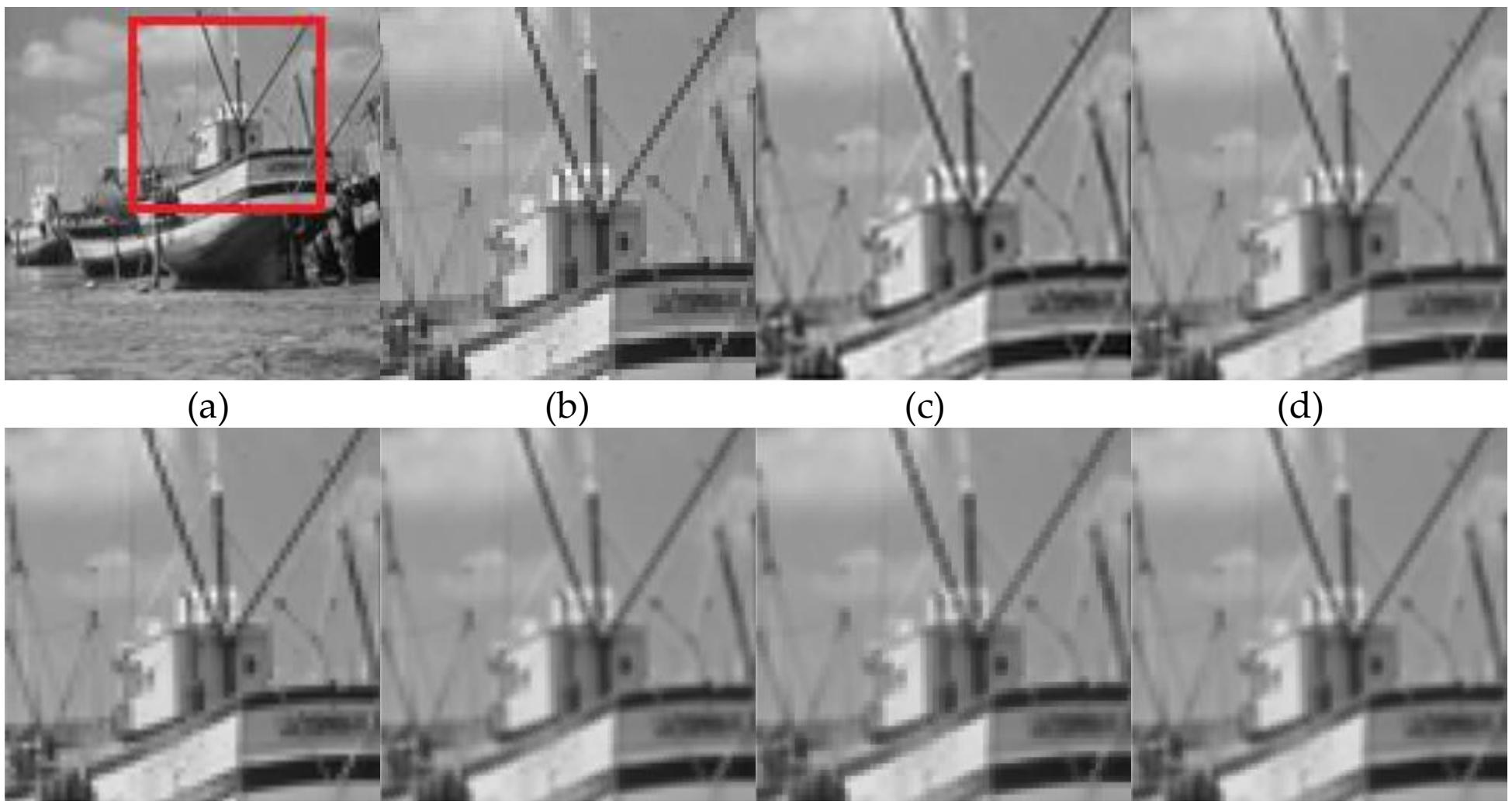

(a) (b) (c) (d)

(e) (f) (g) (h)

**Figure 13.** (a) Boat test image inside a red square: (b) traditional nearest neighbor (TN), (c) traditional bicubic (TC), (d) traditional bilinear (TB), (e) normalized minimum side-based diameter-based (MD), (f) normalized hypotenuse radius-based (HR), (g) normalized virtual pixel length-based height for the area of the triangle-based (AT), and (h) normalized virtual pixel length for hypotenuse-based radius for the area of the circle (AC) interpolation algorithms.

In the MD image, the edges are slightly blurred and jagged. In the HR image, the edges are blurred and slightly jagged. In the AT image, the edges are blurred and jagged, especially at the oblique edges.

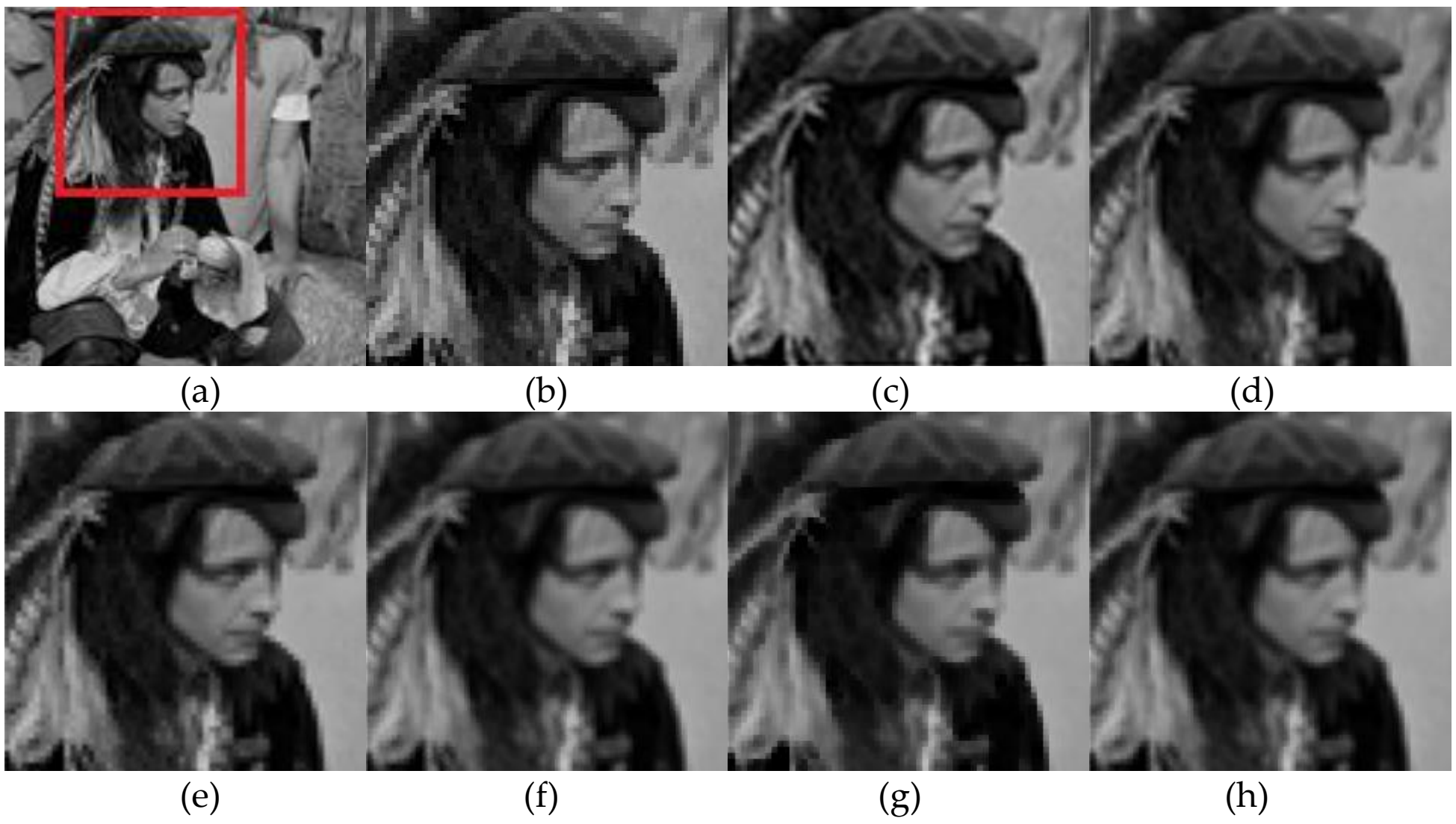

**Figure 14.** (a) Man test image inside a red square: (b) traditional nearest neighbor (TN), (c) traditional bicubic (TC), (d) traditional bilinear (TB), (e) normalized minimum side-based diameter-based (MD), (f) normalized hypotenuse radius-based (HR), (g) normalized virtual pixel length-based height for the area of the triangle-based (AT), and (h) normalized virtual pixel length for hypotenuse-based radius for the area of the circle (AC) interpolation algorithms.

In the AC image, the edges are slightly blurred and slightly jagged, especially at the oblique edges. This may suggest that AC is the best among new schemes and has a visual quality slightly superior to or comparable to that of the TC. This situation is quasi-repeated in Figure 14 and Figure 15, which involved the remaining representative images. It is important to note that the above discussions are observer-dependent. Other observers may reach different conclusions since the subjective evaluations may be influenced by other external factors, for example, viewing distance, the quality of the computer screen, the room lighting condition, and the status of the observer's eyes. It is important to note that the quality of interpolated images may also be affected negatively, to some extent, by compression artefacts (in the attempt to meet the paper format size requirement).

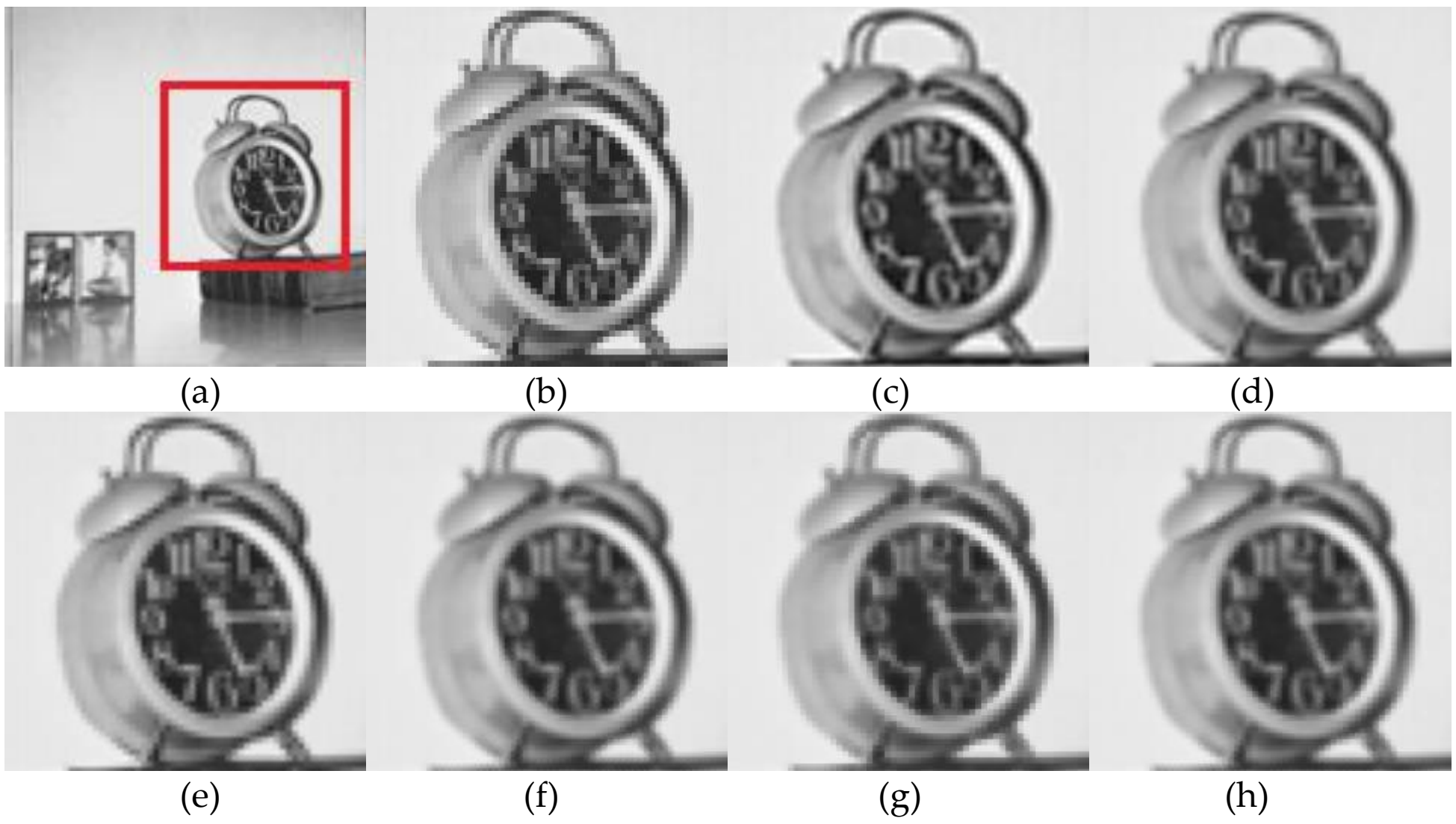

**Figure 15.** (a) Clock test image inside a red square: (b) traditional nearest neighbor (TN), (c) traditional bicubic (TC), (d) traditional bilinear (TB), (e) normalized minimum side-based diameter-based (MD), (f) normalized hypotenuse radius-based (HR), (g) normalized virtual pixel length-based height for the area of the triangle-based (AT), and (h) normalized virtual pixel length for hypotenuse-based radius for the area of the circle (AC) interpolation algorithms.

## 4. Conclusions

Our experiments demonstrated that the introduced four normalized weighting schemes for linear image interpolation algorithms positively affected the image interpolation quality differently depending on the scaling ratio as well as the input image pixels' spatial distribution. The way the normalized weighting schemes affected image interpolation quality was extensively demonstrated objectively (using FR-IQA metrics) and subjectively (by the human observer). To ease the interpretation of the FR-IQA metrics-based results, a simplified interpretation scheme of FR-IQ-based results was also introduced and discussed. It is important to note that, in the first experimental part aimed at comparing all algorithms (i.e., traditional and new scheme methods), when the scaling ratio = 2 and 4, objective image interpolation quality assessment showed that the TC algorithm scored the highest at 100% among all other image interpolation algorithms mentioned. In the second part of our experimental demonstrations, which aimed at comparing new weighting schemes, when the scaling ratio = 2, objective image interpolation quality assessment showed that the MD scored the highest at 0%, HR scored the highest at 66.6%, AT scored the highest at 33.3%, AC scored the highest at 33.3%. In the same part, when the scaling ratio = 4, the objective image interpolation quality assessment showed that the MD scored the highest at 33.3%, HR scored the highest at 0%, AT scored the highest at 0%, and AC scored the highest at 66.6%. Subjective image quality evaluations showed that the overall performance of the AC algorithm was superior or comparable to TB, TC, MD, HR, and AT. Note that, in digital image interpolation, to date, there exists no image interpolation algorithm that best works for all types of digital images. It is also important to note that many medical/biomedical image processing and analysis software still rely on TN, TB, and TC. In terms of computational efficiency, the TN remained the fastest image interpolation algorithm among all algorithms mentioned. Future efforts may focus on developing intelligent weighting schemes capable of producing or generating realistic estimates for image interpolation operations.

**Supplementary Materials:** The following are available online at https://github.com/orukundo, Images size = 512 × 512, 256 × 256, and 128 × 128.

**Funding:** This research work was supported by Lund University between July and December 2020

**Institutional Review Board Statement:** Not applicable.

**Informed Consent Statement:** Not applicable

**Data Availability Statement:** Data are available on our servers and can be shared upon request.

**Conflicts of Interest:** The authors declare no conflict of interest.

## References

1. Wang, Y.; Elhag, T. On the normalization of interval and fuzzy weights. *Fuzzy Sets Syst.* **2006**, *157*, 2456–2471.
2. Pavlacka, O. On various approaches to normalization of interval and fuzzy weights. *Fuzzy Sets Syst.* **2014**, *243*, 110–130.
3. González, R.; Woods, R. *Digital Image Processing*; Pearson; 3rd edition; 2007; Waltham Abbey, United Kingdom; p. 85.
4. Daniel, F.; Julia, K. A *Student's Guide to the Mathematics of Astronomy*; Cambridge University Press: 2013; p. 35.
5. Sally, J.; Sally, P. Chapter 3: Pythagorean triples. In *Roots to Research: A Vertical Development of Mathematical Problems*; American Mathematical Society Bookstore: 2007; p. 63.

6. Sadiq, A.; Almohammad, T.; Khadri, R.A.; Ahmed, A.A.; Lloret, J. An Energy-Efficient Cross-Layer approach for cloud wireless green communications. In Proceedings of the 2017 Second International Conference on Fog and Mobile Edge Computing (FMEC), Valencia, Spain, 8–11 May 2017; pp. 230–234.
7. Fu, H.; Yang, L.; Zhou, C. A computer-aided geometric approach to inverse kinematics. *J. Robot. Syst.* **1998**, *15*, 131–143.
8. Rukundo, O. *Optimal Methods Research on Grayscale Image Interpolation*; China National Knowledge Infrastructure CNKI, TP391.41; Beijing; China; 2012.
9. Sheppard, W. Interpolation. In *Encyclopædia Britannica. 14*, 11th ed.; Chisholm, H., Ed.; Cambridge University Press: 1911; pp. 706–710.
10. Rukundo, O. Evaluation of Rounding Functions in Nearest-Neighbour Interpolation. *Int. J. Comput. Methods* **2021**, *18*, 2150024.
11. Rukundo, O. Effects of Image Size on Deep Learning. *ArXiv* **2021**, arXiv:2101.11508
12. Tian, Q.C.; Wen, H.; Zhou, C.; Chen, W. A fast edge-directed interpolation algorithm. In *International Conference on Neural Information Processing, LNCS*; Huang, T.W., Zeng, Z.G., Li, C.D., Lueng, C.S., Eds.; Springer, Berlin, Heidelberg; 2012; Volume 7665, pp. 398–405.
13. Khan, S.; Lee, D.; Khan, M.A.; Siddiqui, M.F.; Zafar, R.F.; Memon, K.H.; Mujtaba, G. Image Interpolation via Gradient Correlation-Based Edge Direction Estimation. *Sci. Program.* **2020**, *2020*, 5763837.
14. Huang, Z.; Cao, L. Bicubic interpolation and extrapolation iteration method for high resolution digital holographic reconstruction. *Opt. Lasers Eng.* **2020**, *130*, 106090.
15. Lee, Y.; Yu, N.; Tsai, C. An image-upscaling engine for 1080p to 4k using gradient-based interpolation. *Int. J. Electron.* **2020**, *107*, 1386–1405.
16. Xu, G.; Ling, R.; Deng, L.; Wu, Q.; Ma, W. Image interpolation via gaussian-sinc interpolators with partition of unity. *Computers. Mater. Contin.* **2020**, *62*, 309–319.
17. Zulkifli, N.A.B.; Karim, S.A.A.; Shafie, A.B.; Sarfraz, M.; Ghaffar, A.; Nisar, K.S. Image Interpolation Using a Rational Bi-Cubic Ball. *Mathematics* **2019**, *7*, 1045.
18. Rukundo, O.; Wu, K.; Cao, H. Image Interpolation Based on The Pixel Value Corresponding to The Smallest Absolute Difference. In Proceedings of the 4th International Workshop on Advanced Computational Intelligence, Wuhan, China, 19–21 October 2011; pp. 434–437.
19. Rukundo, O.; Maharaj, B. Optimization of Image Interpolation based on Nearest Neighbour Algorithm. In Proceedings of the 9th International Conference on Computer Vision Theory and Applications (VISAPP 2014), Lisbon, Portugal, 5–8 January 2014, pp. 641–647.
20. Rukundo, O.; Pedersen, M.; Hovde, Ø. Advanced Image Enhancement Method for Distant Vessels and Structures in Capsule Endoscopy. *Comput. Math. Methods Med.* **2017**, *2017*, 9813165.
21. Rukundo, O.; Schmidt, S. Aliasing Artefact Index for Image Interpolation Quality Assessment. In Proceedings of the SPIE 10817, Optoelectronic Imaging and Multimedia Technology V, Beijing, China, 7 November 2018; Volume 108171E.
22. Rukundo, O. Half-Unit Weighted Bilinear Algorithm for Image Contrast Enhancement in Capsule Endoscopy. In Proceedings of the SPIE 10615, Ninth International Conference on Graphic and Image Processing (ICGIP 2017), Qingdao, China, 10 April 2018; Volume 106152Q.
23. Rukundo, O.; Schmidt, E.; Ramm, O. Software Implementation of Optimized Bicubic Interpolated Scan Conversion in Echocardiography. *arXiv* **2020**, 1–10, arXiv:2005.11269.
24. Rukundo, O. Effects of Empty Bins on Image Upscaling in Capsule Endoscopy. In Proceedings of the SPIE 10420, Ninth International Conference on Digital Image Processing (ICDIP 2017), Hong Kong, China, 21 July 2017; Volume 104202P.
25. Rucka, M.; Wojtczak, E.; Zielińska, M. Interpolation methods in GPR tomographic imaging of linear and volume anomalies for cultural heritage diagnostics. *Measurement* **2020**, *154*, 107494.
26. Chen, Y; Sun, W.; Li, L.; Chang, C.; Wang, X. An efficient general data hiding scheme based on image interpolation. *J. Inf. Secur. Appl.* **2020**, *54*, 102584.
27. Wang, X.; Jia, X.; Zhou, W.; Qin, X.; Guo, H. Correction for color artifacts using the RGB intersection and the weighted bilinear interpolation. *Appl. Opt.* **2019**, *58*, 8083–8091.
28. Hassan, F.; Gutub, A. Efficient reversible data hiding multimedia technique based on smart image interpolation. *Multimed. Tools Appl.* **2020**, *79*, 30087–30109.
29. Jiang, C.; Li, H.; Zhou, S.; Yu, J.; Chen, L.; Xie, X. Image interpolation model based on packet losing network. *Multimed. Tools Appl.* **2020**, *79*, 25785–25800.
30. De Feis, I.; Masiello, G.; Cersosimo, A. Optimal Interpolation for Infrared Products from Hyperspectral Satellite Imagers and Sounders. *Sensors* **2020**, *20*, 2352.
31. Moraes, T.; Amorim, P.; Da Silva, J.V.; Pedrini, H. Medical image interpolation based on 3D Lanczos filtering. *Comput. Methods Biomech. Biomed. Eng. Imaging Vis.* **2020**, *8*, 294–300.
32. Huang, W.; Liu, J. Robust Seismic Image Interpolation with Mathematical Morphological Constraint. *IEEE Trans. Image Process.* **2020**, *29*, 819–829.
33. Song, G.; Qin, C.; Zhang, K.; Yao, X.; Bao, F.; Zhang, Y. Adaptive Interpolation Scheme for Image Magnification Based on Local Fractal Analysis. *IEEE Access* **2020**, *8*, 34326–34338.
34. Murad, M.; Bilal, M.; Jalil, A.; Ali, A.; Mehmood, K.; Khan, B. Efficient Reconstruction Technique for Multi-Slice CS-MRI Using Novel Interpolation and 2D Sampling Scheme. *IEEE Access* **2020**, *8*, 117452–117466.

35. Ji, J.; Zhong, B.; Ma, K. Image Interpolation Using Multi-Scale Attention-Aware Inception Network. *IEEE Trans. Image Process.* **2020**, *29*, 9413–9428.
36. Chung, K.; Chen, S. An effective bilinear interpolation-based iterative chroma subsampling method for color images. *Multimed Tools Appl* **2022**, *81*, 32191–32213.
37. Yu, L.; Liu, K.; Orchard, M.T. Orchard, Manifold-Inspired Single Image Interpolation. *arXiv* **2021**, arXiv:2108.00145.
38. Gao, C.; Zhou, R.; Li, X. Quantum color image scaling based on bilinear interpolation. *Chin. Phys. B* **2022**., doi: 10.1088/1674-1056/ac8f35
39. Sadaghiani, A.; Sheikhaei, S.; Forouzandeh, B. Image Interpolation Based on 2D-DWT with Novel Regularity-Preserving Algorithm Using RLS Adaptive Filters. *Int. J. Image Graph.* **2022**.
40. Occorsio, D.; Ramella, G.; Themistoclakis, W. Image Scaling by de la Vallée-Poussin Filtered Interpolation. *J. Math. Imaging Vis.* **2022**.
41. Zhou, H.; Xu, Z.; Tian, Y.; Yu, Z.; Zhang, Y.; Ma, J. Interpolation-based nonrigid deformation estimation under manifold regularization constraint. *Pattern Recognit.* **2022**, *128695;* doi: 10.1016/j.patcog.2022.108695
42. Fei, Y.; Shan, Z.; Salvador, E.; Kaoru, K.H. Implementing bilinear interpolation with quantum images. *Digit. Signal Process.* **2021**, *117*, 103149.
43. Tavoosi, J.; Zhang, C.; Mohammadzadeh, A.; Mobayen, S.; Mosavi, A. Medical Image Interpolation Using Recurrent Type-2 Fuzzy Neural Network. *Front. Neuroinform. Volume 15, 667375;* **2021.**
44. Romano, Y.; Isidoro, J.; Milanfar, P. RAISR: Rapid and Accurate Image Super Resolution. IEEE Trans. Comput. Imaging 2017, 3, 110–125.
45. Dong, C.; Loy, C.; He, K.; Tang, X. Learning a Deep Convolutional Network for Image Super-Resolution. In FComputer Vision, ECCV; Leet D., Pajdla, T., Schiele, B., Tuytelaars, T., Eds.; Volume 8692. Springer, Cham, 2014.
46. Rukundo, O. Non-extra Pixel Interpolation. *Int. J. Image Graph.* **2020**, *20*, 2050031.
47. Rukundo, O.; Schmidt, S. Stochastic Rounding for Image Interpolation and Scan Conversion. Int. J. Adv. Comput. Sci. Appl. 2022, 13, 13–22.
48. Rukundo, O.; Schmidt, S. Effects of Rescaling Bilinear Interpolant on Image Interpolation Quality. In Proceedings of the SPIE 10817, Optoelectronic Imaging and Multimedia Technology V, Beijing, China, 2 November 2018; Volume 1081715.
49. Rukundo, O.; Schmidt, S. Extrapolation for Image Interpolation. In Proceedings of the SPIE 10817, Optoelectronic Imaging and Multimedia Technology V, Beijing, China, 2 November 2018; Volume 108171F.
50. Zhang, L.; Wu, X. An edge-guided image interpolation algorithm via directional filtering and data fusion. *IEEE Trans. Image Process.* **2006**, *15*, 2226–2238.
51. Li, X.; Orchard, M.T. Orchard: New edge-directed interpolation. *IEEE Trans. Image Process.* **2001**, *10*, 1521–1527.
52. Rukundo, O.; Huang, M.; Cao, H. Optimization of Bilinear Interpolation Based on Ant Colony Algorithm. In Proceedings of the 2nd International Conference Electrical and Electronics Engineering, Macao, China, 1–2 December 2011; pp. 571–580.
53. Rukundo, O.; Cao, H. *Advances on Image Interpolation Based on Ant Colony Algorithm*; SpringerPlus: Berlin/Heidelberg, Germany, 2016; Volume 5, p. 403.
54. Rukundo, O.; Cao, H. Nearest Neighbor Value Interpolation. *Int. J. Adv. Comput. Sci. Appl. (IJACSA)* **2012**, *3*, 25–30.
55. Rukundo, O. Effects of Improved-Floor Function on the Accuracy of Bilinear Interpolation Algorithm. *Comput. Inf. Sci.* **2015**, *8*, 1–25.
56. Mittag, U.; Kriechbaumer, A.; Rittweger, J. A novel interpolation approach for the generation of 3D-geometric digital bone models from image stacks. J. Musculoskelet. Neuronal Interact. 2017, 17, 86–96.
57. Wang, Y.; Zhang, Z.; Guo, B. 3D image interpolation based on directional coherence. In Proceedings IEEE Workshop on Mathematical Methods in Biomedical Image Analysis (MMBIA 2001), Kauai, HI, USA, 9–10 December 2001; pp. 195–202.
58. Quadrilaterals. Available online: https://www.mathsisfun.com/quadrilaterals.html (accessed on 1 November 2020).
59. List of Geometry and Trigonometry Symbols, Math Vault. Available online: https://mathvault.ca/hub/higher-math/math-symbols/ geometry-trigonometry-symbols/ (accessed on 1 November 2020).
60. USC-SIPI Image Database. Available online: http://sipi.usc.edu/database/database.php (accessed on 8 November 2020).
61. Modified-USC-SIPI-Image-Database. Available online: https://github.com/orukundo/Modified-USC-SIPI-Image-Database (accessed on 8 November 2020).
62. Zhou; Bovik, W.A.C.; Sheikh, H.R.; Simoncelli, E.P. Image Quality Assessment: From Error Visibility to Structural Similarity. *IEEE Trans. Image Process.* **2004**, *13*, 600–612.